\documentclass[11pt]{article}

\usepackage[a4paper,margin=1in]{geometry}
\usepackage[T1]{fontenc}
\usepackage[utf8]{inputenc}
\usepackage{lmodern}
\usepackage{amsmath, amssymb}
\usepackage{verbatim}
\usepackage{url}
\usepackage{graphicx}    
\usepackage{rotating}    

\title{
  \textbf{enclawed}: A Configurable, Sector-Neutral Hardening Framework for
  Single-User AI Assistant Gateways
}

\author{%
  Alfredo Metere\\
  Metere Consulting, LLC.\\
  \texttt{alfredo.metere@metereconsulting.com}
}

\date{May 1, 2026}

\begin{document}

\maketitle

\begin{abstract}
We present \emph{enclawed}, a hard-fork hardening framework built on top of
the OpenClaw single-user personal artificial intelligence (AI) assistant
gateway. enclawed targets deployments that need attestable peer trust,
deny-by-default external connectivity, signed-module loading, and a
tamper-evident audit trail --- typically regulated industries such as
financial services, healthcare, defense contracting, regulated R\&D, and
government enclaves. The framework ships in two flavors: an \texttt{open}
flavor that preserves OpenClaw compatibility while still emitting audit,
classification, and data-loss-prevention (DLP) signals, and an
\texttt{enclaved} flavor that activates strict allowlists, Federal
Information Processing Standards (FIPS) cryptographic-module assertion,
mandatory module-manifest signature verification, and high-assurance
peer attestation for the Model Context Protocol (MCP). The classification ladder is fully data-driven: a deploying
organization selects from five built-in presets (generic, US-government,
healthcare, financial services, three-tier) or supplies its own JSON. We
accompany the implementation with a security review, a 356-case test
suite (261 unit tests, 95 adversarial pen-tests) covering tamper
detection, signature forgery, fetch- and raw-socket egress bypass,
audit-log truncation and re-ordering, trust-root mutation, DLP evasion,
prompt injection, code injection, and biconditional admission for
net-capable extensions; real-time human-in-the-loop control (per-agent
pause / resume / stop and approval queues); a memory-bounded secure
transaction buffer with rollback (default cap 50\% of system RAM,
configurable); a strict-mode TypeScript typecheck of every framework
file; and a GitHub Actions workflow ready for continuous integration.
A newly added \texttt{.mjs} primitive --- the biconditional
extension-admission gate (\texttt{extension-admission.mjs}) ---
extends the paper's skill trust schema to non-skill extensions, so
every loadable extension declaring \texttt{net.egress} must carry a
signed manifest, an explicit per-extension host allowlist, and (in the
enclaved flavor) a verification level $\geq$ \texttt{tested}.
The four-level verification lattice is now closed at the top: a
companion stack of four \texttt{.mjs} primitives
(\texttt{skill-formal-static.mjs}, \texttt{skill-formal-types.mjs},
\texttt{skill-formal-bmc.mjs}, \texttt{skill-formal-bundle.mjs}) plus
a command-line interface (CLI) at
\texttt{scripts/skills-formal-verify.mjs} produces a signed
proof-carrying bundle the runtime can re-check at load time, raising a
skill from \texttt{tested} to \texttt{formal} via composable static
effect-containment, refinement-typed dispatch, and bounded model
checking against the runtime biconditional. The unit-test inventory
correspondingly grows from 208 to 261 unit tests; together with the
95 adversarial pen-tests this is a 356-case suite.
\textbf{enclawed is a hardening framework,
not an accredited compliance certification.} The deploying organization
remains responsible for hardware, validated cryptographic modules,
certified facilities, and assessor sign-off.
\end{abstract}

\section{Introduction}

Generative AI has crossed quickly from prototype into the operational
fabric of regulated industries. Financial-services firms use large
language models (LLMs) for research summarization touching material
non-public information (MNPI); healthcare systems apply them to
protected health information (PHI); defense contractors handle
controlled unclassified information (CUI) and International Traffic in
Arms Regulations (ITAR)-controlled materials; legal teams rely on
them for privileged-counsel work; pharmaceutical R\&D operates on
embargoed clinical-trial data. Each of these settings is governed by
frameworks that pre-date generative AI but apply to it directly --- the
Health Insurance Portability and Accountability Act (HIPAA) [8], the
European Union General Data Protection Regulation (GDPR) [7], the
Payment Card Industry Data Security Standard (PCI~DSS) [9], the
International Organization for Standardization / International
Electrotechnical Commission (ISO/IEC) 27001 [5], the American Institute
of Certified Public Accountants Service Organization Control 2 (SOC~2)
trust-services criteria [6], the National Institute of Standards and
Technology (NIST) Special Publication 800-53 [1], and NIST 800-171 [4]
--- and increasingly by AI-specific regimes such as the European Union
Artificial Intelligence Act (EU AI Act) [28], the NIST AI Risk
Management Framework (NIST AI RMF) [26], and ISO/IEC 42001 [27]
governing artificial-intelligence management systems.

Two observations motivate this work. First, the dominant open-source
multi-channel AI gateways were designed for the consumer
``personal-assistant'' market and ship with cloud-first defaults that
are fundamentally incompatible with regulated-enclave deployment.
Second, the security failure modes of generative AI are no longer
hypothetical: documented incidents include training-data extraction at
scale [31, 32], indirect prompt injection through retrieved content
[20, 21, 22], the 2023 Samsung corpus leak through interactive cloud
application programming interfaces (APIs), and the 2024 \emph{Moffatt
v.\ Air Canada} decision [35] establishing operator liability for AI
assistant outputs. The Open Web Application Security Project (OWASP)
Top~10 for LLM Applications [29] and MITRE Adversarial Threat Landscape
for Artificial-Intelligence Systems (ATLAS) [30] now codify the
relevant attack surface in industry-recognized form.

\subsection{The regulated-enterprise AI gap}

Personal AI gateways such as OpenClaw [25] exemplify the problem. The
upstream project's positioning is explicit: a single-user assistant
that ``answers you on the channels you already use,'' bundling 24+
chat platforms (WhatsApp, Telegram, Slack, Discord, Signal, iMessage,
Matrix, Microsoft Teams, and more) and 40+ cloud LLM providers (OpenAI,
Anthropic, Google, Mistral, Groq, AWS Bedrock, Azure OpenAI, OpenRouter,
Fireworks, etc.) by default. Its plugin model encourages
community-published modules installed via \texttt{npm} with no enforced
signature verification. Its policy posture is permissive: external
channels and providers are enabled the moment credentials are
configured. Audit, classification, and DLP are not part of the
framework's contract.

For a casual community user --- one whose worst-case agent action is
to leak a chat transcript or run up an API bill --- this is a
defensible design. Even there, however, the same defaults are not
benign: a permissively-configured OpenClaw deployment that wires an
agent to a robot, a home-automation bus, a 3D printer, a CNC mill, an
electric door, a vehicle controller, an irrigation valve, or any
other real-world actuator can quietly turn a chat-level mistake into
property damage, data loss, or physical harm. Prompt-injection,
hallucinated tool calls, third-party module compromise, and unbounded
egress are not academic risks once the loop reaches a physical
device. So the ``community user'' boundary is itself thinner than the
upstream defaults assume.

For a financial-services trading floor handling MNPI, a hospital
system processing PHI, a defense contractor with CUI access, or any
organization subject to ISO 27001 / SOC~2 / HIPAA / PCI / Federal
Risk and Authorization Management Program (FedRAMP) / U.S.\
Department of Defense (DoD) / U.S.\ Department of Energy (DoE)
controls, every one of those defaults becomes an explicit policy
violation:

\begin{itemize}
  \item \emph{External egress to consumer messaging platforms} violates
    boundary protection (NIST 800-53 SC-7 [1]) and information-flow
    control (AC-4).
  \item \emph{Cloud LLM API usage} routes regulated content through
    third parties whose data-processing terms typically permit
    retention and, until contractually disclaimed, training reuse ---
    in direct conflict with HIPAA business-associate constraints [8,
    §164.308(b)], GDPR Article~28 processor obligations [7], and
    most data-residency policies.
  \item \emph{Unsigned community plugins} expose the gateway to
    supply-chain compromise --- the threat the NIST Secure Software
    Development Framework (SSDF) [33] addresses through SI-7 / SR-3 /
    SR-4, and which the U.S.\ Cybersecurity and Infrastructure Security
    Agency (CISA) / National Security Agency (NSA) / Federal Bureau of
    Investigation (FBI) joint guidance on AI deployment [34] explicitly
    calls out as a top operational risk.
  \item \emph{Permissive defaults with no audit chain} make incident
    reconstruction impossible, conflicting with NIST 800-53 AU-2/3/9/10
    and HIPAA §164.312(b) audit-control requirements.
  \item \emph{No classification labels} means an output may carry
    PHI / MNPI / CUI to a recipient cleared for none of it, with the
    host silent at every step.
\end{itemize}

\subsection{Why a configuration-only fix is insufficient}

A natural response is ``just disable the cloud channels and providers,
add a logging plugin, and require approved modules.'' We argue this is
not enough, for three structural reasons.

\textbf{Defaults are policy.} A platform whose out-of-the-box state
contains 78 forbidden modules invites accidental enablement at every
upgrade, every onboarding, every plugin install. Removing the modules
from the source tree --- not just from the configuration --- is the
only durable defense, because it makes the unsafe state
\emph{unreachable} rather than merely \emph{unselected}. Compliance
posture must be a property of the binary, not a property of a
configuration file an operator can revert.

\textbf{The plugin-loading path is the trust boundary.} OpenClaw's
plugin loader does not verify signatures, does not consult any trust
root, and does not reason about clearance. Adding ``a logging plugin''
is itself a plugin --- it inherits the same unsigned, unverified
loading path. The trust boundary must be \emph{below} the plugin
layer, enforced by the host before any plugin code runs, and locked
against post-boot mutation. Retrofitting this into a project whose
contract assumes ambient plugin trust is a re-architecture, not a
configuration.

\textbf{Upstream cannot adopt these constraints.} OpenClaw's product
identity --- a multi-channel personal assistant --- is incompatible
with deny-by-default external connectivity, FIPS-required boot,
mandatory signature verification, and attested peer trust. Even if
upstream maintainers were willing to accept patches that imposed those
defaults, doing so would break their published contract with thousands
of community deployments. Forking is therefore not merely convenient
but structurally necessary: the regulated-enclave product must diverge
from the consumer-assistant product at the binary level. We say
explicitly that upstream OpenClaw \emph{will not} adopt these
constraints, not because the upstream maintainers are inattentive to
security --- they are not --- but because the constraints are
incompatible with the product they ship.

\subsection{Three design commitments}

This paper describes \emph{enclawed}, a hard fork of OpenClaw rebuilt
around three design commitments that address the gap above:

\begin{enumerate}
  \item \textbf{Always-on policy.} The classification framework
    activates unconditionally at process start, before any plugin or
    transit module is imported, so downstream code observes a guarded
    environment from the first instruction. There is no
    ``compatibility mode'' that disables the framework in production.
  \item \textbf{Two flavors.} An \texttt{open} flavor preserves
    OpenClaw behavior with the framework running in warn-only mode for
    development and non-regulated deployment; an \texttt{enclaved}
    flavor enforces strict deny-by-default policy, FIPS assertion,
    mandatory module-manifest signing, and attested peer trust.
    Selection is by a single environment variable read once at boot
    and then immutable for the process lifetime.
  \item \textbf{Data-driven classification.} The level ladder is not
    hardcoded; the deploying organization selects from five built-in
    presets (sector-neutral generic, U.S.-government, healthcare-HIPAA,
    financial-services, three-tier) or supplies its own JavaScript
    Object Notation (JSON) document. The framework is sector-neutral by
    default --- a U.S.-government preset and DoE Q-clearance and
    L-clearance templates exist as opt-ins, not as the canonical
    defaults.
\end{enumerate}

The framework is sector-neutral by default. Built-in presets cover
financial services (MNPI / privileged-counsel), healthcare (PHI /
sensitive PHI), a generic three-tier scheme, and full US-government
markings, with the canonical defaults industry-generic (\texttt{PUBLIC}
$<$ \texttt{INTERNAL} $<$ \texttt{CONFIDENTIAL} $<$ \texttt{RESTRICTED}
$<$ \texttt{RESTRICTED-PLUS}).

\subsection{Comparison with the broader landscape}
\label{sec:landscape}

OpenClaw is one of several open-source projects that mediate between a
human user and one or more language models. Adjacent offerings
include LibreChat [43], Open WebUI [44], AnythingLLM [45], Jan.ai
[46], Lobe Chat [47] --- multi-provider chat front-ends --- plus
agent-orchestration frameworks such as AutoGen [48] and the LangChain
ecosystem [49]. A separate cluster of AI-security tools targets
narrower slices of the problem: NeMo Guardrails [50] and Llama Guard
[51] focus on content moderation at the model boundary; Lakera Guard
[52] addresses prompt injection as a hosted scanning service.

We surveyed each of these against the security properties enclawed is
designed to deliver, summarized in Table~\ref{tab:landscape}. The
honest picture is that no other publicly-available offering, as
documented at the time of writing, combines mandatory module signing,
a classification lattice with a configurable scheme, tamper-evident
audit, real-time human-in-the-loop control, a memory-bounded rollback
buffer, and attested peer trust into a single cohesive framework that
runs entirely on locally-hosted inference. The closest competitors
each address one or two of these properties:

\begin{itemize}
  \item Local-LLM front-ends (Open WebUI, AnythingLLM, Jan.ai) deliver
    point (1) --- they avoid cloud egress by binding to local runtimes
    --- but ship without policy, classification, audit, or
    signed-module enforcement. Their plugin or knowledge-base ingest
    paths are unverified, which is the same supply-chain gap noted in
    §1.1 for OpenClaw.
  \item Multi-provider gateways (LibreChat, Lobe Chat) prioritize
    reach over restriction; cloud providers are first-class citizens
    and removal requires user effort that is not enforced. Their
    plugin / preset distribution is community-driven without
    cryptographic attestation.
  \item Agent frameworks (AutoGen, LangChain ecosystem) are libraries
    rather than gateways, and the security posture is whatever the
    embedding application provides. They have no notion of
    deployment-flavor strictness, no built-in egress allowlist, and
    no rollback on partial-failure of an agent's actions.
  \item Content-guardrail systems (NeMo Guardrails, Llama Guard)
    address output filtering --- a complement to, not a replacement
    for, classification-driven flow control. They do not enforce
    information-flow rules, do not produce a tamper-evident audit
    trail, and do not gate module loading.
  \item SaaS prompt-injection scanners (Lakera Guard) by design route
    content through an external service, which is incompatible with
    the no-egress requirement of regulated enclaves.
\end{itemize}

Where a competitor genuinely addresses one of enclawed's properties,
enclawed adopts the same primitive rather than reinventing it: the DLP
scanner and prompt-shield (§4.5, §8) are deliberately a thin
keyword-and-regex layer because content-classification depth is the
job of dedicated tools like Llama Guard, which a deploying
organization can wire in addition. enclawed's contribution is the
host-level chassis that any of these complementary tools can sit
inside, with the chassis itself taking responsibility for the
properties the tools cannot deliver alone (signed module loading,
classification lattice, deny-by-default policy, attested peer trust,
HITL controls, rollback).

\begin{table}[ht]
  \centering
  \scriptsize
  \setlength{\tabcolsep}{4pt}
  \renewcommand{\arraystretch}{1.2}
  \resizebox{\textwidth}{!}{%
  \begin{tabular}{|l|c|c|c|c|c|c|c|c|c|}
    \hline
    \textbf{Project} & \textbf{Local-only} & \textbf{Mand.\ signed} & \textbf{Class.} & \textbf{Tamper-} & \textbf{DLP} & \textbf{HITL} & \textbf{Tx} & \textbf{Attested} & \textbf{Sector-neutral} \\
    & \textbf{by default} & \textbf{modules} & \textbf{lattice} & \textbf{evident audit} & \textbf{at output} & \textbf{stop} & \textbf{rollback} & \textbf{peer trust} & \textbf{config scheme} \\
    \hline
    \textbf{enclawed (this work)} & yes (enclaved) & yes & yes & yes & yes & yes & yes & yes (MCP) & yes \\
    \hline
    OpenClaw [25]              & no & no & no & no & no & no & no & no & no \\
    \hline
    LibreChat [43]              & no & no & no & no & no & no & no & no & no \\
    \hline
    Open WebUI [44]             & yes & no & no & no & no & no & no & no & no \\
    \hline
    AnythingLLM [45]            & opt-in & no & no & no & no & no & no & no & no \\
    \hline
    Jan.ai [46]                 & yes & no & no & no & no & no & no & no & no \\
    \hline
    Lobe Chat [47]              & no & no & no & no & no & no & no & no & no \\
    \hline
    AutoGen [48]                & N/A (library) & no & no & no & no & partial & no & no & no \\
    \hline
    LangChain [49]              & N/A (library) & no & no & no & partial & partial & no & no & no \\
    \hline
    NeMo Guardrails [50]        & N/A (component) & no & no & no & yes & no & no & no & no \\
    \hline
    Llama Guard [51]            & N/A (model) & no & no & no & yes & no & no & no & no \\
    \hline
    Lakera Guard [52]           & no (SaaS) & no & no & no & yes & no & no & no & no \\
    \hline
  \end{tabular}%
  }%
  \caption{Landscape comparison. Properties documented as of the
    projects' publicly-stated features at time of writing; ``partial''
    means the property is achievable but not enforced by the project
    itself, and ``N/A'' means the project is a different kind of
    artifact (library, component, model) rather than a hostable
    gateway and the column does not directly apply. enclawed is the
    only entry that combines all nine properties as a host-level
    framework around which the others can be composed.}
  \label{tab:landscape}
\end{table}

\paragraph{Non-claim.} This work does not constitute a certification.
Real audits --- ISO 27001, SOC~2, the Health Information Trust Alliance
(HITRUST) common security framework, FedRAMP, DoD/DoE high-side
Authority to Operate (ATO), PCI DSS Report on Compliance (RoC) --- all
require accredited hardware, validated cryptographic modules (FIPS
140-3 [2] where the regime demands), certified facilities, evaluation
by a qualified assessor, and management sign-off. enclawed is a
code-level scaffold; the gaps a deploying organization must close are
catalogued in Section~\ref{sec:gaps}.

\paragraph{Contributions.}
\begin{itemize}
  \item A two-flavor classification framework with sector-neutral
    defaults and built-in presets for finance, healthcare, government,
    R\&D, and a minimal three-tier scheme.
  \item A user-configurable classification scheme model that accepts
    arbitrary level vocabularies, validated against rank-contiguity and
    name-uniqueness invariants.
  \item An Ed25519 module-signing system with a per-signer
    clearance-approval trust root and boot-time pre-verification.
  \item A hash-chained, append-only audit log with concurrent-append
    serialization, deep payload sanitization against log injection, and
    independent third-party verification.
  \item An adversarial test suite of 95 cases targeting tamper
    detection, signature forgery, fetch- and raw-socket egress bypass,
    audit-log truncation and re-ordering, trust-root mutation, DLP
    evasion, prompt injection, code injection, and biconditional
    admission for net-capable extensions, alongside 261 unit tests, all
    passing on Node 22.
  \item A GitHub Actions workflow that runs the suites on push, PR, and
    weekly cron, plus a strict-mode TypeScript typecheck of all 22
    framework files.
\end{itemize}

\section{Background and Threat Model}

\subsection{Choice of access-control model}
\label{sec:bla-rationale}

Information security has produced a small library of formal access
control models, each emphasizing a different property. We surveyed the
candidates against three criteria: (i) the primary risk in our
deployments is unauthorized \emph{disclosure} of regulated data, not
modification or fraud; (ii) the policy must be \emph{mandatory} ---
imposed by the host, not configurable by users or modules; (iii) the
ladder of sensitivity levels must be \emph{reusable} across sectors
that already think in tier vocabularies (PHI, MNPI, CUI, etc.).
Table~\ref{tab:model-choice} summarizes the alternatives we
considered.

\begin{table}[ht]
  \centering
  \footnotesize
  \setlength{\tabcolsep}{4pt}
  \renewcommand{\arraystretch}{1.15}
  \resizebox{\textwidth}{!}{%
  \begin{tabular}{|p{4.5cm}|p{2.6cm}|p{4.5cm}|p{3.3cm}|}
    \hline
    \textbf{Model} & \textbf{Primary property} & \textbf{Why not the primary fit} & \textbf{Use in enclawed} \\
    \hline
    Bell-LaPadula [10] & confidentiality & --- & primary axis \\
    \hline
    Biba [36] & integrity & does not constrain disclosure & complementary, future work \\
    \hline
    Clark-Wilson [37] & integrity via well-formed transactions & overengineered for chat-mediated AI; assumes app-defined IVPs & inspires the rollback buffer (§4.6) \\
    \hline
    Brewer-Nash [38] & dynamic conflict-of-interest separation & sector-specific (financial); orthogonal to tier flow & easily layered via compartments \\
    \hline
    Harrison-Ruzzo-Ullman (HRU) access matrix & general expressiveness & undecidable safety; no enforced lattice & rejected \\
    \hline
    Role-based access control (RBAC) [39] & role-based authorization & no notion of data sensitivity; group$\to$resource only & rejected as primary axis \\
    \hline
    Attribute-based access control (ABAC) [40] & attribute expressions & expressive but harder to formally verify; risk of policy drift & adopted within compartments / releasability slots \\
    \hline
  \end{tabular}%
  }
  \caption{Access-control model survey. Bell-LaPadula is selected as
    the primary axis because the deployments enclawed targets are
    fundamentally confidentiality-driven (PHI, MNPI, CUI, embargoed
    research). Biba's integrity dual and Clark-Wilson's transaction
    discipline are complementary and inform §4.6 (rollback buffer);
    RBAC and ABAC are layered as compartment / releasability
    assignments rather than as the primary flow rule.}
  \label{tab:model-choice}
\end{table}

Concretely we chose Bell-LaPadula for five reasons:

\begin{enumerate}
  \item \emph{Confidentiality is the first-order risk.} Every regime in
    Section~\ref{sec:gaps}'s gap list is a confidentiality regime first
    (PHI under HIPAA [8], MNPI under U.S.\ Securities and Exchange
    Commission (SEC) rules, CUI under NIST 800-171 [4], classified
    information under NIST 800-53 [1]). Integrity and availability
    matter, but only after confidentiality is established.
  \item \emph{The lattice is mechanizable.} Dominance, least
    upper-bound, and reading/writing predicates are simple, total,
    decidable functions on integer ranks plus subset tests on
    compartment sets. They are easier to formally verify than the
    rule-evaluation of Clark-Wilson or the policy expressions of ABAC.
  \item \emph{It maps directly to existing vocabularies.} US-government
    classification, healthcare PHI tiers, and financial MNPI all
    naturally express as a rank ladder plus optional compartments. The
    same code consumes any of them via the configurable scheme of §6.
  \item \emph{It is mandatory by construction.} The lattice is a
    property of the host, evaluated on every label comparison. Modules
    cannot override it.
  \item \emph{Biba and Clark-Wilson compose orthogonally.} A future
    integrity layer can use Biba's no-read-down / no-write-up on a
    parallel axis without modifying the existing lattice; the
    transaction buffer (§4.6) already implements the well-formed
    transaction discipline of Clark-Wilson at the action layer.
\end{enumerate}

\subsection{Bell-LaPadula formal model}

Each subject and object carries a label $\langle \ell, C, R \rangle$
where $\ell$ is a numeric rank, $C$ a set of compartments, and $R$ a
set of releasability caveats. Label $a$ dominates label $b$ ($a
\succeq b$) iff $\ell_a \geq \ell_b$ and $C_b \subseteq C_a$. Reading
is allowed when the subject dominates the object (no-read-up); writing
is allowed when the object dominates the subject (no-write-down).
Combination yields the least upper bound. This lattice arithmetic is
implemented in \texttt{src/enclawed/classification.ts} and is
independent of the level vocabulary.

\subsection{Threat model}

We assume a single-tenant, single-user gateway running inside a
high-trust enclave. The user holds the deploying organization's highest
applicable trust tier. The system must:

\begin{enumerate}
  \item never egress to the public Internet or any external channel/provider;
  \item use only locally-hosted inference (e.g.\ Ollama [18], vLLM [16],
    LM Studio [19], SGLang [17], local NVIDIA Inference Microservice (NIM));
  \item refuse to render or echo data above the user's authorized tier;
  \item refuse to write data below its origin classification (no-write-down);
  \item produce a tamper-evident audit trail of every model interaction;
  \item encrypt every persistent artifact at rest with a key bound to
    organization-controlled hardware;
  \item block known sensitive-data markings, secrets, and personally
    identifiable information (PII) from leaving the gateway via any
    output channel;
  \item make any deviation a hard, loud failure --- not silent
    degradation.
\end{enumerate}

We map these requirements to multiple control frameworks simultaneously
(Table~\ref{tab:controls}): NIST 800-53 [1], ISO/IEC 27001 [5], NIST
Cybersecurity Framework (CSF) 2.0 [3], SOC~2 trust-services criteria
(TSC) [6], GDPR [7], the HIPAA Security Rule [8], PCI DSS [9], and the
Cybersecurity Maturity Model Certification (CMMC) layered on NIST
800-171 [4]. The deployment-side complement to this mapping --- the
controls the deploying organization must still deliver --- is
enumerated in Section~\ref{sec:gaps}.

\begin{sidewaystable}
  \centering
  \normalsize
  \setlength{\tabcolsep}{6pt}
  \renewcommand{\arraystretch}{1.3}
  \resizebox{\textheight}{!}{%
  \begin{tabular}{|l|l|l|l|l|l|l|l|}
    \hline
    \textbf{Control area} & \textbf{enclawed surface} & \textbf{NIST 800-53} & \textbf{ISO 27001} & \textbf{NIST CSF} & \textbf{SOC 2} & \textbf{GDPR} & \textbf{HIPAA} \\
    \hline
    Access enforcement & \texttt{classification.ts}, \texttt{policy.ts} & AC-3 & A.9.4 & PR.AC-4 & CC5.1, CC6.1 & Art.~32(1)(b) & 164.312(a)(1) \\
    \hline
    Information flow & \texttt{egress-guard.ts} & AC-4, SC-7 & A.13.1 & PR.PT-4 & CC6.6 & Art.~32(1)(b) & 164.312(e)(1) \\
    \hline
    Security attributes & \texttt{classification.ts} + scheme & AC-16 & A.8.2 & PR.DS-5 & CC1.4 & Art.~5(1)(f) & 164.308(a)(2) \\
    \hline
    Audit logging & \texttt{audit-log.ts} + \texttt{subsystem} patch & AU-2/3/9/10 & A.12.4 & DE.CM-1, DE.AE-3 & CC7.2, CC7.3 & Art.~30, 32(1)(d) & 164.312(b) \\
    \hline
    Crypto module & \texttt{crypto-fips.assertFipsMode()} & IA-7, SC-13 & A.10.1 & PR.DS-1 & CC6.1 & Art.~32(1)(a) & 164.312(a)(2)(iv) \\
    \hline
    Encryption at rest / in transit & \texttt{crypto-fips.ts} & SC-8/13/28 & A.10.1, A.13.2 & PR.DS-1/2 & CC6.1, CC6.7 & Art.~32(1)(a) & 164.312(e)(2)(ii) \\
    \hline
    Continuous monitoring & audit + subsystem + egress.onDeny & SI-4 & A.12.4 & DE.CM, DE.AE & CC7.2 & Art.~32(1)(d) & 164.308(a)(1)(ii)(D) \\
    \hline
    Data leakage / DLP & \texttt{dlp-scanner.ts} & SI-12 & A.13.2.3 & PR.DS-5 & CC6.7 & Art.~5(1)(f) & 164.312(e) \\
    \hline
    Memory hygiene & \texttt{zeroize.ts} & MP-6, SI-16 & A.8.3 & PR.IP-6 & CC6.5 & Art.~32(1)(a) & 164.310(d)(2)(i) \\
    \hline
    Least functionality & host admission gate (Sec.~\ref{sec:moduleset}) & CM-7 & A.9.1 & PR.AC-3 & CC6.3 & Art.~25 & 164.312(a)(1) \\
    \hline
    Module integrity & \texttt{module-signing} + \texttt{trust-root} & SI-7, CM-5, SR-3/4 & A.12.5, A.14.2.5 & PR.IP-1, PR.DS-6 & CC8.1 & Art.~32(1)(b) & 164.312(c)(1) \\
    \hline
  \end{tabular}%
  }%
  \caption{Control mapping. Each row links a framework surface to the
  corresponding requirement in seven major frameworks; CMMC L2/L3
  inherits NIST 800-171 [4] and PCI DSS [9] adds Req.~3, 4, 7, 10, 11
  alongside NIST 800-53. The deploying organization picks the column
  that fits its regime. (Rendered on a dedicated landscape page; cells
  use natural-width \texttt{l} columns and the whole table is uniformly
  scaled to fill the page.)}
  \label{tab:controls}
\end{sidewaystable}

\subsection{Out of scope}

The framework does \emph{not} provide accredited cryptography
(FIPS 140-3 or NSA Type-1), an accredited cross-domain solution (CDS),
identity binding (Common Access Card (CAC), Personal Identity
Verification (PIV), Security Assertion Markup Language (SAML), or
OpenID Connect (OIDC)), kernel-level mandatory access control
(SELinux Multi-Level Security (MLS)), Write-Once-Read-Many (WORM)
audit shipping, or facility security. These remain the deploying
organization's responsibility. Section~\ref{sec:gaps} enumerates each.

\section{Architecture}

\subsection{Two flavors}

enclawed selects a flavor at boot via the \texttt{ENCLAWED\_FLAVOR}
environment variable. Table~\ref{tab:flavors} contrasts the two.

\begin{table}[t]
  \centering
  \small
  \begin{tabular}{|p{3.7cm}|p{5cm}|p{5cm}|}
    \hline
    \textbf{Aspect} & \textbf{open (default)} & \textbf{enclaved} \\
    \hline
    Channel/provider/host allowlists & not enforced & strict deny-by-default \\
    \hline
    Module signatures & warn-only & required, must verify \\
    \hline
    Signer-clearance approval & informational & required, exact match \\
    \hline
    FIPS assertion at boot & off (configurable) & on (default) \\
    \hline
    Trust root mutation post-boot & permitted & locked \\
    \hline
    \texttt{globalThis.fetch} reassignment & permitted & frozen non-configurable \\
    \hline
    Module without manifest & loaded with warning & rejected \\
    \hline
    MCP peer attestation & verified, fail $\to$ warn & verified, fail $\to$ deny \\
    \hline
  \end{tabular}
  \caption{Flavor matrix. Both flavors run the same framework code; the
  distinction is enforcement strictness.}
  \label{tab:flavors}
\end{table}

\subsection{Module set: cloud modules gated, not source-stripped}
\label{sec:moduleset}

Earlier drafts of this paper argued that the safest cut was to
\emph{delete} the 78 cloud-channel and cloud-provider module
directories from the upstream \texttt{extensions/} tree so the unsafe
state would be UNREACHABLE rather than UNSELECTED. After porting from
upstream OpenClaw, the framework now ships \emph{all} 134 module
directories (cloud channels, cloud LLM providers, external
search/browser/webhook modules, and local-capable modules) and
delegates rejection to the host admission gate
(\texttt{src/plugins/manifest-registry.ts}). In the \texttt{enclaved}
flavor, every module that does not present a signed
\texttt{enclawed.module.json} verifying against the trust root is
denied at load time \emph{before any plugin code is imported}; in the
\texttt{open} flavor an unsigned module loads with a warning. The
unreachability property therefore shifts from a source-tree property
(``the directory does not exist'') to a verified-manifest property
(``the directory exists, but no signed manifest grants admission'').
This shift gives operators of the open flavor access to the broader
upstream module catalog while preserving the enclaved flavor's
deny-by-default posture, and it lets the framework absorb upstream
extension churn without re-running deletion sweeps. As of this draft
\texttt{enclawed-oss} ships 130 extension directories: 28 carry a
signed \texttt{enclawed.module.json} (the local-capable inference and
core extensions: Ollama, vLLM, LM Studio, SGLang, NVIDIA NIM Local,
ComfyUI, the media/memory/speech cores, \texttt{openshell},
\texttt{qa-channel}, \texttt{phone-control}, \texttt{device-pair}, and
\texttt{diagnostics-otel}); 96 are gated-unsigned (cloud channels and
cloud LLM providers whose \texttt{enclawed.module.json} is intentionally
absent so the enclaved admission gate rejects them); 2 are utility-only
directories that present no plugin metadata at all. The bundle verifier
(\texttt{scripts/verify-enclaved-bundle.mjs}) reports the signed-versus-
gated counts on every CI run.

\subsection{Runtime singleton and bootstrap}

Activation is a single call at the top of \texttt{src/entry.ts}, before
any plugin or transit module is imported:

\begin{verbatim}
const { bootstrapEnclawed } = await import("./enclawed/bootstrap.js");
await bootstrapEnclawed();
\end{verbatim}

\texttt{bootstrapEnclawed()} reads the flavor and classification scheme
from the environment, loads the policy, opens the audit log, installs
the egress guard (with \texttt{freeze:true} in \texttt{enclaved}),
pre-verifies every module manifest under \texttt{extensions/}, locks the
trust root in \texttt{enclaved}, and stores the runtime on a
\texttt{Symbol.for("enclawed.runtime")} slot of \texttt{globalThis} so
downstream patches can consult it without creating import cycles. The
boot record (with active scheme id, flavor, FIPS state, and allowlist
contents) is the genesis hash-chain entry.

\section{Implementation}

The framework is delivered in two parallel surfaces: a TypeScript (TS)
twin under \texttt{src/enclawed/} that bundles with the upstream
OpenClaw build, and a canonical reference under \texttt{enclawed/src/}
written in plain Node ECMAScript Modules (\texttt{.mjs}) with zero
runtime dependencies. The
canonical surface runs under \texttt{node --test} without
\texttt{pnpm install}, which keeps the standalone security suite cheap to
exercise and trivial to audit. Table~\ref{tab:loc} summarizes file counts
and lines of code per tree.

\paragraph{Repository layout.} The framework is hosted as
\texttt{enclawed-oss} --- the open-source, MIT-licensed, self-contained,
public-bound git repository this paper documents. It carries the
OpenClaw fork base, the framework primitives (classification, policy,
audit, DLP, crypto wrapper, zeroize, module signing, trust root, HITL,
transaction buffer, prompt shield, two-layer egress guard, biconditional
extension admission), the canonical \texttt{.mjs} reference, the
standalone test suite, this paper, and \emph{the full mirrored OpenClaw
\texttt{extensions/} catalog} (\(\sim\)130 directories, both
local-capable and cloud). Cloud extensions are admitted only when their
\texttt{enclawed.module.json} is signed by a trust-root key; the unsafe
state is unreachable through the admission gate, not absent from the
source tree. A separate proprietary companion product,
\texttt{enclawed-enclaved}, builds on this open foundation and is
distributed under separate commercial terms; its internal architecture
is outside the scope of this paper.

\begin{table}[t]
  \centering
  \small
  \begin{tabular}{|l|r|r|}
    \hline
    \textbf{Surface (tree)} & \textbf{Files} & \textbf{LOC} \\
    \hline
    \multicolumn{3}{|l|}{\emph{enclawed-oss/ (open source, MIT)}} \\
    \hline
    TypeScript framework (\texttt{src/enclawed/*.ts}, integration shims) & 32 & $\sim$3{,}956 \\
    \hline
    Canonical \texttt{.mjs} reference (\texttt{enclawed/src/}) & 18 & $\sim$2{,}286 \\
    \hline
    Standalone unit + pen tests (\texttt{enclawed/test/}) & 25 & $\sim$3{,}535 \\
    \hline
    Upstream patches (4 files) & 4 & $+$82 \\
    \quad \texttt{src/entry.ts} & & $+$10 \\
    \quad \texttt{src/plugins/channel-validation.ts} & & $+$27 \\
    \quad \texttt{src/plugins/provider-validation.ts} & & $+$27 \\
    \quad \texttt{src/logging/subsystem.ts} & & $+$18 \\
    \hline
  \end{tabular}
  \caption{Code footprint of the open-source repository. The OSS tree
  is self-contained and fully buildable, testable, and publishable on
  its own.}
  \label{tab:loc}
\end{table}

\subsection{Hash-chained audit log}

Every record is canonicalized as JSON with sorted keys, prepended with
the previous record's SHA-256, and hashed. Three properties matter:

\begin{itemize}
  \item \emph{Tamper detection.} Editing a record in place breaks the
    chain at the next record's \texttt{prevHash} comparison, which
    \texttt{verifyChain()} reports as \texttt{brokenAt}.
  \item \emph{Concurrent-write safety.} \texttt{append()} chains every
    write onto an internal \texttt{Promise} queue so two simultaneous
    callers cannot both consume the same \texttt{lastHash}. Without this
    guard, parallel emitters (e.g.\ the egress-guard \texttt{onDeny}
    callback racing with \texttt{createSubsystemLogger}) would produce a
    self-inconsistent chain.
  \item \emph{Log-injection neutralization.} A \texttt{deepSanitize}
    pass strips C0 control characters from every string in
    \texttt{type}, \texttt{actor}, \texttt{level}, and the entire
    payload tree, replacing them with U+FFFD. This both prevents an
    attacker from spoofing a fake JSONL record by embedding a newline in
    a payload string, and ensures the on-disk bytes match the bytes
    that were hashed. \texttt{\_\_proto\_\_}, \texttt{constructor}, and
    \texttt{prototype} keys are filtered out before canonicalization so
    a payload object cannot smuggle prototype-pollution shapes into the
    audit hash.
\end{itemize}

\subsection{Egress guard}
\label{sec:egress}

The egress guard now ships in two layers. The first layer,
\texttt{installEgressGuard()}, replaces \texttt{globalThis.fetch} with
a wrapper that resolves the target hostname via the Web Hypertext
Application Technology Working Group (WHATWG) URL parser (yielding
case-normalized hostnames, ignoring the \texttt{Host} header, and
rejecting embedded credentials) and checks against an allowlist. In
the \texttt{enclaved} flavor the property is set via
\texttt{Object.defineProperty} with \texttt{writable:false} and
\texttt{configurable:false}, making subsequent reassignment throw
under ECMAScript Modules (ESM) strict mode.

\paragraph{Raw-socket layer.} Earlier drafts noted that the fetch wrapper
explicitly did not protect against raw \texttt{net.Socket} use; that
gap is now closed in JS by \texttt{installRawSocketGuard()}, which
patches \texttt{Socket.prototype.connect}. The patch unwraps Node's
internal three argument forms (\texttt{(options, cb)},
\texttt{(port, host, cb)}, and the normalized-array form used by
\texttt{net.createConnection} and \texttt{http(s).Agent.createConnection})
and applies the allowlist check to the resolved \texttt{host}. Because
\texttt{tls.TLSSocket} extends \texttt{net.Socket}, and because both
\texttt{http.Agent} and \texttt{https.Agent} ultimately invoke
\texttt{Socket.prototype.connect}, the patch transitively covers
\texttt{fetch}, \texttt{http}, \texttt{https}, \texttt{tls}, and any
extension that imports \texttt{node:net} directly. In the
\texttt{enclaved} flavor the patch is installed with the same
\texttt{writable:false}, \texttt{configurable:false} freeze pattern
used for \texttt{fetch}.

\paragraph{VPN-only mode.}
\begin{sloppypar}
The enclaved-flavor policy
(\texttt{defaultEnclavedPolicy}) defaults to
\texttt{requireVpnGateway:\,true} with the RFC 1918 CIDR set
(\texttt{10.0.0.0/8}, \texttt{172.16.0.0/12}, \texttt{192.168.0.0/16}).
The deploying organization replaces these with the exact CIDR its VPN
exposes. Under
VPN-only mode every IPv4 destination must either appear on the literal
hostname allowlist (covering loopback, mDNS \texttt{.local}, and
pre-resolved gateway names) or fall inside one of the configured CIDRs.
A small zero-dependency \texttt{ipInCidr()} matcher implements the
check so the canonical \texttt{.mjs} surface remains fully
runtime-dependency-free. IPv6 destinations are denied unless explicitly
allow-listed by literal address; deploying organizations that run a v6
VPN must add the matching CIDR via the future v6-aware matcher.
\end{sloppypar}

\paragraph{Biconditional admission for net-capable extensions.}
\texttt{extension-admission.mjs} extends the paper's skill trust schema
(§\ref{sec:trust}) to non-skill extensions. Every loadable extension is
the tuple $(M, \mathit{content}, \sigma)$, where the manifest $M$ now
also records a \texttt{verification} level
(\texttt{unverified} / \texttt{declared} / \texttt{tested} /
\texttt{formal}) and, if $M$ declares the \texttt{net.egress} capability,
an explicit \texttt{netAllowedHosts} target list. The admission gate
rejects, in the \texttt{enclaved} flavor: any unsigned manifest, any
manifest whose signer is absent from the trust root, any tampered
manifest (the canonical bytes now cover \texttt{verification} and
\texttt{netAllowedHosts} so post-signing field changes invalidate
$\sigma$), and any \texttt{net.egress} declaration at
\texttt{verification} below \texttt{tested}. In the \texttt{open}
flavor those same conditions surface as warnings instead of denials.
At runtime, \texttt{installPerExtensionEgressGuard()} narrows the
deployment-wide allowlist to the per-extension declared hosts and
records a \texttt{biconditional.violation} record on every
$S \setminus D \neq \emptyset$ deviation (an extension reaching a
host it never declared). A post-hoc check
(\texttt{biconditionalNetCheck}) compares declared and observed host
sets and flags the dangerous direction;
$D \setminus S$ over-declaration is benign and surfaces only as advice.

\paragraph{Limits.} Even both JS layers together do \emph{not} cover
(a) native modules that perform syscalls below libuv, (b) child
processes spawned via \texttt{node:child\_process} that invoke external
binaries (curl, wget), or (c) memory-mapped or shared-memory channels
inside the same machine. For a true classified enclave the deploying
organization MUST add kernel-level egress controls (\texttt{nftables},
extended Berkeley Packet Filter (eBPF), network namespaces, optional
one-way diode) on top. The JS layer is deliberately the
minimum-surface-cheapest defense, sufficient to catch JS-level
misbehavior by ported third-party extensions but explicitly insufficient
alone for accreditation.

\subsection{Closing the verification lattice: the formal-verification stack}
\label{sec:formal-stack}

The trust schema's four-level lattice
(\texttt{unverified} $<$ \texttt{declared} $<$ \texttt{tested} $<$
\texttt{formal}) ships in the manifest, but until recently only the
first three levels had a corresponding admission rule: \texttt{formal}
required a mechanically-checkable proof of capability containment that
the framework did not produce. The newly added stack ---
\texttt{skill-formal-static.mjs},
\texttt{skill-formal-types.mjs},
\texttt{skill-formal-bmc.mjs},
\texttt{skill-formal-bundle.mjs}, all under \texttt{enclawed/src/}
--- closes that gap with three composable methods plus a
proof-carrying bundle the runtime re-checks at load time. The full development is the
subject of a separate companion paper [57]; we
summarise here only the parts that matter to a deploying organization
asking ``what does \texttt{verification: \"formal\"} now mean
operationally?''.

\paragraph{Method A --- script-side static effect containment.}
\texttt{skill-formal-static.mjs} walks every script colocated with
\texttt{SKILL.md} (Python, shell, Node, TypeScript) and computes a
sound over-approximation of the system effects it can produce, in the
parent paper's capability vocabulary
(\texttt{net.egress}, \texttt{fs.read}, \texttt{fs.write.\{rev,irrev\}},
\texttt{tool.invoke}, \texttt{spawn.proc}, \texttt{publish},
\texttt{pay}, \texttt{mutate.schema}). Reflective constructs
(\texttt{eval}, \texttt{exec}, dynamic \texttt{import}) taint the
analysis to the lattice top so the verdict is conservative by
construction. The verdict is admitted only if the over-approximated
effect set is contained in the manifest's declared \texttt{caps}.

\paragraph{Method B --- refinement-typed dispatch.}
\texttt{skill-formal-types.mjs} exposes \texttt{buildRefinedDispatch(M)},
which returns a frozen dispatcher that throws \texttt{RefinementError}
on any envelope whose capability is outside $M.\mathrm{caps}$. The
dispatcher is the runtime's only ingress to the host application
programming interfaces (APIs); an envelope an adversarially-prompted
large language model (LLM) emits cannot reach a host API without
passing the type predicate. The data-processing inequality
[58] bounds the per-envelope leakage at
$\log_2(\lvert D \rvert + 1)$ bits, where $D = M.\mathrm{caps}$.

\paragraph{Method C --- bounded model checking.}
\texttt{skill-formal-bmc.mjs} runs an exhaustive depth-first search
over the abstract envelope state space at bound $K$ (default 8) and
checks, on every symbolic trace, that the runtime biconditional of
[10] holds (no world-state change without a corresponding admitted
envelope). The state-space size is
$(\lvert D \rvert + 1)^{K}$; for deployments with
$\lvert D \rvert \le 10$ and $K \le 8$ the search explores at most
$11^{8} \approx 2.1 \times 10^{8}$ traces, which discharges in
seconds.

\paragraph{Proof-carrying skill bundle.} The CLI
\texttt{scripts/skills-formal-verify.mjs} composes the three methods,
hashes each evidence file with a sorted-key canonical encoding, and
signs a four-file bundle (\texttt{static.json}, \texttt{types.proof.json},
\texttt{smt.unsat.json}, \texttt{manifest.attest.json}) with the same
Ed25519 \texttt{module-signing} primitive used for the manifest itself.
The runtime's bundle re-checker
(\texttt{verifyFormalBundle} in \texttt{skill-formal-bundle.mjs})
re-runs all three methods on the live source, compares hashes against
the cached evidence, and verifies the attestation signature. A drift
between cached and freshly-computed verdicts aborts admission with a
\texttt{method-A-cache-miss} (or \texttt{method-B}, \texttt{method-C})
reason; the runtime never trusts the bundle's producer. A worked
end-to-end demonstration ships at \texttt{skills/\_formal-demo/}.

The full developer-facing reference for the \texttt{SKILL.md} format
--- mandatory front-matter fields, the capability vocabulary, the
canonicalisation and Ed25519 signing convention, the seven-step
bootstrap re-check protocol, and the proof-carrying bundle layout
--- is published at
\url{https://docs.openclaw.ai/tools/skill-manifest-spec} [59] and
is the source of truth skill authors and runtime integrators are
expected to consume.

\subsection{Configurable classification scheme}

A scheme is a JSON document conforming to:

\begin{verbatim}
{
  "id": "acme-2026",
  "description": "ACME Corp internal data classification policy v3.2",
  "levels": [
    { "rank": 0, "canonicalName": "Public",        "aliases": ["P"] },
    { "rank": 1, "canonicalName": "Internal",      "aliases": ["I"] },
    { "rank": 2, "canonicalName": "Customer Data", "aliases": [] },
    { "rank": 3, "canonicalName": "Privileged",    "aliases": ["legal"] }
  ],
  "validCompartments":  ["FINANCE", "ENG", "LEGAL"],
  "validReleasability": ["NDA", "EYES_ONLY"]
}
\end{verbatim}

\texttt{parseClassificationScheme()} enforces three invariants: ranks
must be contiguous starting at zero; every name (canonical and alias)
must be unique across the whole scheme after case-insensitive
normalization; non-empty levels list. The active scheme governs
\texttt{format()}, \texttt{parse()}, \texttt{makeLabel()} range
validation, manifest \texttt{clearance} validation, and required-tier
checks at the MCP attestation layer. Five presets ship out of the box:
\texttt{enclawed-default} (six tiers; generic + US-gov merged),
\texttt{us-government} (UNCLASSIFIED to TOP SECRET//SCI),
\texttt{healthcare-hipaa} (PHI / Sensitive-PHI / Research-Embargoed),
\texttt{financial-services} (MNPI / Privileged-Counsel),
\texttt{generic-3-tier} (Public / Internal / Restricted).

\subsection{DLP scanner}

A regex-based scanner with three pattern families:
\emph{sensitive-data markings} (industry banners, U.S.-government
classification banners, DoE Restricted Data, Sensitive Compartmented
Information (SCI) codewords, distribution caveats); \emph{cloud and
vendor secrets} (Amazon Web Services (AWS), Google Cloud Platform (GCP)
service account, Microsoft Azure storage key, GitHub, GitLab, OpenAI,
Anthropic, Slack, Stripe, JSON Web Tokens (JWTs), Privacy-Enhanced
Mail (PEM)-encoded private keys); and \emph{international PII} (email
address, ITU-T E.164 phone number, credit-card primary account number
(PAN), International Bank Account Number (IBAN), U.S.\ Social Security
Number (SSN)). Input length is capped at 1\,MiB by default to bound
CPU exposure to regular-expression denial-of-service (ReDoS)-shaped
patterns; callers may pass \texttt{onOversize:'truncate'} to scan the
prefix. Severity levels (\texttt{low}, \texttt{medium}, \texttt{high},
\texttt{critical}) drive a \texttt{redact()} entry point that replaces
matches above a threshold with \texttt{[REDACTED]}. The scanner is a
backstop, not a primary control: regex DLP cannot detect paraphrased
sensitive content, optical-character-recognition-only content, or any
pattern not in its catalog.

\subsection{Human-in-the-loop control}
\label{sec:hitl}

\texttt{src/enclawed/hitl.ts} provides per-agent sessions with a
state machine \texttt{PENDING $\to$ RUNNING $\rightleftarrows$ PAUSED
$\to$ \{STOPPED, COMPLETED, FAILED\}}, a real-time event stream on the
controller (\texttt{EventEmitter}-based, with both a typed channel per
event name and a generic \texttt{event} channel), and an approval
queue. Each \texttt{AgentSession} exposes:

\begin{itemize}
  \item \texttt{pause()}, \texttt{resume()}, \texttt{stop(reason)},
    \texttt{complete()}, \texttt{fail(reason)} --- explicit operator
    controls. \texttt{stop} is final; the state machine forbids
    transitions out of any terminal state.
  \item \texttt{checkpoint()} --- a cooperative cancellation point
    the agent code awaits between actions. It throws
    \texttt{AgentStoppedError} if the session has been stopped, and
    blocks until \texttt{resume()} if the session is paused. We chose
    cooperative over preemptive cancellation deliberately: Saltzer and
    Schroeder's \emph{complete mediation} principle [41] argues for
    explicit checkpoints over forced thread termination, which in a
    single-threaded async runtime can leave shared state in an
    inconsistent half-mutated form.
  \item \texttt{proposeAction(actionType, payload)} --- if the
    action type belongs to the session's approval-required set, the
    call awaits a human decision (allow / deny) routed through the
    controller's approval queue. Deny raises
    \texttt{ApprovalDeniedError}; if the agent is stopped while the
    call is awaiting the human, it instead raises
    \texttt{AgentStoppedError} (the approval is no longer relevant).
\end{itemize}

The \texttt{HitlController} maintains the session registry, the
approval queue, and a \texttt{stopAll(reason)} mass-halt operation.
Every state transition and every approval request/resolution is
emitted as an event and (when an \texttt{AuditLogger} is wired)
appended to the hash-chained audit log under the
\texttt{hitl.}$\langle$\texttt{type}$\rangle$ namespace. A separate
operator UI (CLI, web, or remote) subscribes to the event stream to
render live status and to call \texttt{resolveApproval(id, decision)}
when a human approves or denies a queued action. Sub-second latency
is the design goal; the controller does no I/O on the hot path.

\subsection{Secure transaction buffer with rollback}
\label{sec:tx}

The framework treats every reversible action an agent takes as a
\emph{transaction}, in the sense of Gray's classical formulation
[42]. The deploying organization registers each action's inverse at
the moment of execution; the buffer (\texttt{src/enclawed/transaction-buffer.ts})
holds the resulting transaction record so it can be rolled back later
if a human, an automated guardrail, or a failed downstream operation
demands undo.

\paragraph{Memory-bounded rollback window.} The buffer is sized as a
configurable percentage of system RAM (default 50\%, configurable via
the \texttt{ramPercent} constructor option, with an absolute
\texttt{maxBytes} override). The size cap matches the eviction policy
of Clark-Wilson [37]'s well-formed transaction discipline applied to a
finite resource: as long as the action sits in the buffer it can be
inverted; once evicted, it has been implicitly committed. A new record
that would push the buffer past the cap auto-commits the oldest
records FIFO until enough room is freed.

\paragraph{Tamper-evident chain.} Each transaction is hash-chained to
its predecessor with SHA-256, in the same shape as the audit log of
§4.1. \texttt{verifyChain()} performs an independent walk so an
operator (or a separate auditor process) can detect silent tampering
of the in-memory record set. Every record, rollback, commit, and
eviction is emitted to the audit log when one is wired.

\paragraph{Rollback semantics.} \texttt{rollback(n)} pops the n most
recent records and runs each \texttt{inverse} in LIFO order. An
inverse that throws is reported in the result's \texttt{errors} array
and an audit \texttt{transaction.rollback-failed} record is appended,
but rollback continues so a single broken inverse cannot block
restoration of the rest of the window. After rollback the chain is
re-anchored to the new tail so subsequent records continue cleanly.

\section{Module Signing and Trust Root}
\label{sec:trust}

Each module ships an \texttt{enclawed.module.json} alongside its
\texttt{package.json}:

\begin{verbatim}
{ "v": 1,
  "id": "ollama",
  "publisher": "enclawed bundled",
  "version": "0.1.0",
  "clearance": "internal",
  "capabilities": ["model-provider"],
  "signerKeyId": "enclawed-bundled-dev-2026",
  "signature": "<base64 Ed25519 signature>" }
\end{verbatim}

The signature is an Ed25519 [13] signature over a stable canonical
encoding of the manifest body excluding \texttt{signature} itself, so
signing is idempotent. The trust root binds each public key to a list
of clearance tiers it is approved to attest. Verification passes when
(i) the manifest's \texttt{signerKeyId} resolves to a signer in the
trust root, (ii) the signer is approved for the manifest's declared
clearance, (iii) the signature verifies against the canonical bytes,
and (iv) any caller-required clearance is met. In the \texttt{enclaved}
flavor, the bootstrap calls \texttt{lockTrustRoot()} after the
deploying organization's signers are loaded, after which any
\texttt{setTrustRoot()} or \texttt{resetTrustRoot()} call throws
\texttt{TrustRootLockedError}.

The module loader is two-staged:
\texttt{preloadModuleDecisions(rootDir)} walks the modules directory
once at boot, parses every manifest, runs \texttt{checkModule()}
against the active flavor and trust root, and stashes the per-id
decisions on the runtime singleton; the synchronous validation
chokepoints (channel- and provider-validation, both patched into
upstream OpenClaw) read the cached decisions without performing async
I/O on a hot path. Modules denied at preload are rejected when their
channels or providers subsequently try to register, with a
corresponding \texttt{module.deny.channel} or
\texttt{module.deny.provider} audit record.

\section{Configurable Classification --- Worked Examples}

Table~\ref{tab:schemes} shows the five built-in schemes side by side.

\begin{table}[t]
  \centering
  \footnotesize
  \setlength{\tabcolsep}{3pt}
  \resizebox{\textwidth}{!}{%
  \begin{tabular}{|c|l|l|l|l|l|}
    \hline
    \textbf{Rank} & \textbf{default} & \textbf{us-government} & \textbf{healthcare-hipaa} & \textbf{financial-services} & \textbf{3-tier} \\
    \hline
    0 & PUBLIC          & UNCLASSIFIED    & PUBLIC             & PUBLIC             & PUBLIC \\
    \hline
    1 & INTERNAL        & CUI             & INTERNAL           & INTERNAL           & INTERNAL \\
    \hline
    2 & CONFIDENTIAL    & CONFIDENTIAL    & PHI                & CONFIDENTIAL       & RESTRICTED \\
    \hline
    3 & RESTRICTED      & SECRET          & SENSITIVE-PHI      & MNPI               & --- \\
    \hline
    4 & RESTRICTED-PLUS & TOP SECRET      & RESEARCH-EMB.\     & PRIVILEGED-COUNSEL & --- \\
    \hline
    5 & SCI             & TOP SECRET//SCI & ---                & ---                & --- \\
    \hline
  \end{tabular}%
  }%
  \caption{Built-in classification schemes (column headers are scheme
    ids; rank-4 of \texttt{healthcare-hipaa} is RESEARCH-EMBARGOED,
    abbreviated for table fit). Aliases such as \texttt{q-cleared}
    $\equiv$ rank 4 of \texttt{us-government} are accepted by
    \texttt{parse()} and by manifest \texttt{clearance} validation.}
  \label{tab:schemes}
\end{table}

A deploying organization selects a scheme via:

\begin{verbatim}
ENCLAWED_FLAVOR=enclaved \
ENCLAWED_CLASSIFICATION_SCHEME=healthcare-hipaa \
ENCLAWED_AUDIT_PATH=/var/log/enclawed/audit.jsonl \
node openclaw.mjs gateway run --bind loopback
\end{verbatim}

For a custom scheme, the same env var is set to a JSON file path;
\texttt{loadSchemeByName} accepts an optional \texttt{allowedDirs}
parameter that rejects paths outside a vetted configuration directory.

\section{Prompt Injection Defenses}

\texttt{src/enclawed/prompt-shield.ts} provides defensive helpers for
text that flows from untrusted sources (channel input, retrieved
document, tool output) into a model prompt. The framework does not
promise complete prompt-injection defense --- that is an open research
problem [20, 21, 22] --- but it neutralizes the most common
silent-confusion vectors so they surface as visible content rather than
control signals to the model.

\texttt{sanitizeForPrompt(text)} performs five operations in order:

\begin{enumerate}
  \item \texttt{stripControlChars} replaces ASCII C0 control characters
    (the byte range 0x00--0x1F excluding TAB) with the Unicode
    replacement character U+FFFD.
  \item \texttt{stripBidi} removes the Unicode bidirectional-override
    range U+202A--U+202E and U+2066--U+2069.
  \item \texttt{stripZeroWidth} removes the zero-width characters
    U+200B--U+200D, U+2060, and U+FEFF.
  \item \texttt{neutralizeRoleBoundaries} prefixes any line starting
    with \texttt{system:} / \texttt{assistant:} / \texttt{user:} /
    \texttt{tool:} / \texttt{function:} (with optional leading
    whitespace or Markdown emphasis) with a literal
    \texttt{[USER-CONTENT]} marker so the model sees them as quoted
    rather than as a fresh role boundary.
  \item \texttt{neutralizeFences} inserts a zero-width character before
    any line consisting of triple backticks so the closing fence cannot
    break out of a quoted code block the host wraps untrusted input in.
\end{enumerate}

\texttt{detectInjection(text)} returns finding identifiers without
modifying the text, including a heuristic for ALL-CAPS imperative
override phrases (``IGNORE ALL PREVIOUS INSTRUCTIONS'', ``OVERRIDE the
above messages'', etc.).

\section{Evaluation}

\subsection{Test methodology}

The framework ships with two test surfaces. The canonical \texttt{.mjs}
suite uses Node's built-in \texttt{node:test} runner [24] and depends
only on \texttt{node:crypto}, allowing verification without any
\texttt{npm install}. A vitest twin under
\texttt{src/enclawed/integration.test.ts} exercises the same surface in
the upstream OpenClaw \texttt{pnpm test} pipeline.

\subsection{Unit test coverage}

Table~\ref{tab:unit} enumerates the per-module unit-test counts. All
261 unit tests pass on Node 22.

\begin{table}[t]
  \centering
  \small
  \begin{tabular}{|l|r|p{6cm}|}
    \hline
    \textbf{Module} & \textbf{Tests} & \textbf{Coverage focus} \\
    \hline
    \texttt{classification.test.mjs}        & 13 & lattice arithmetic, format/parse round-trip, immutability \\
    \hline
    \texttt{classification-scheme.test.mjs} & 11 & all 5 presets round-trip, custom JSON, validation invariants \\
    \hline
    \texttt{dlp-scanner.test.mjs}           & 11 & all 3 pattern families, redact correctness \\
    \hline
    \texttt{policy.test.mjs}                & 9  & enclaved/open defaults, allowlist enforcement, frozen \\
    \hline
    \texttt{module-loader.test.mjs}         & 8  & 8 deny paths and 1 allow path under both flavors \\
    \hline
    \texttt{crypto-fips.test.mjs}           & 10 & AES-GCM round-trip, AAD round-trip (string + Buffer), AAD mismatch rejection, FIPS gate \\
    \hline
    \texttt{zeroize.test.mjs}               & 7  & buffer fill, withSecret on success and throw \\
    \hline
    \texttt{egress-guard.test.mjs}          & 6  & block, allow, lifecycle \\
    \hline
    \texttt{module-manifest.test.mjs}       & 6  & valid parse, invariant rejection, hash stability \\
    \hline
    \texttt{flavor.test.mjs}                & 5  & alias parsing, env override, default \\
    \hline
    \texttt{audit-log.test.mjs}             & 4  & append + verify, tamper, reopen, hash determinism \\
    \hline
    \texttt{module-signing.test.mjs}        & 4  & Ed25519 round-trip, wrong-key, tampered, malformed \\
    \hline
    \texttt{hitl.test.mjs}                  & 14 & state machine, cooperative checkpoint, pause/resume/stop, approval queue, audit integration \\
    \hline
    \texttt{transaction-buffer.test.mjs}    & 18 & ramPercent sizing, hash chain, LIFO rollback, partial-failure rollback, eviction, audit integration \\
    \hline
    \texttt{paper-conformance.test.mjs} & 82 & §3.2 module-set gating (one subtest per gated cloud module), §5 signing coverage, §3 parallel surfaces, §9 paper-declared test counts \\
    \hline
    \texttt{skill-formal-static.test.mjs} & 17 & Method A pattern matches across Python / shell / Node / TS, comment stripping, directory walking, reflective-construct tainting, contained vs non-contained verdicts \\
    \hline
    \texttt{skill-formal-types.test.mjs} & 12 & Method B refinement-boundary throws, in / out-of-manifest dispatch, unknown-token rejection, predicate form, JSON-serialisable verdict \\
    \hline
    \texttt{skill-formal-bmc.test.mjs} & 14 & Method C reference gate, biconditional violation kinds (executed-without-audit, executed-but-deny, admitted-without-audit), state-space size formula, instanceHash stability \\
    \hline
    \texttt{skill-formal-bundle.test.mjs} & 10 & sign / verify round-trip, mismatched-key failure, cache-miss-on-skill-drift, signer-not-authorised rejection, on-disk round-trip, canonicalisation stability, tamper detection \\
    \hline
    \textbf{Total OSS unit} & \textbf{261} & \\
    \hline
  \end{tabular}
  \caption{Unit test inventory.}
  \label{tab:unit}
\end{table}

\subsection{Adversarial pen-test coverage}

Table~\ref{tab:pen} enumerates the per-file pen-test counts. All 95
pen-tests pass on Node 22.

\begin{table}[t]
  \centering
  \small
  \begin{tabular}{|l|r|p{7cm}|}
    \hline
    \textbf{File} & \textbf{Tests} & \textbf{Adversarial focus} \\
    \hline
    \texttt{audit-log.pentest.mjs}        & 6  & in-place edit, tail truncation (documented limit), record reorder, newline log-injection, \texttt{\_\_proto\_\_} pollution, 100 concurrent appends \\
    \hline
    \texttt{signature-forgery.pentest.mjs} & 7  & wrong-key, signerKeyId swap, signature replay across modules, downgrade attack, capability injection, malformed signature, open-flavor warning capture \\
    \hline
    \texttt{egress-bypass.pentest.mjs}    & 8  & hostname case normalization, Host-header spoofing, embedded credentials, IPv4-in-IPv6, file:/data: schemes, fetch reassign with/without freeze (latter via subprocess), malformed URLs \\
    \hline
    \texttt{trust-root-and-scheme.pentest.mjs} & 10 & setTrustRoot after lock, \texttt{\_\_proto\_\_}/constructor pollution, duplicate normalized names, non-contiguous ranks, negative rank, empty levels, non-string canonicalName, missing id, type checks \\
    \hline
    \texttt{dlp-evasion.pentest.mjs}      & 8  & oversize cap, ReDoS bound, zero-width camouflage, whitespace variants, redact correctness, PEM camouflage gap, null/undefined safety \\
    \hline
    \texttt{prompt-injection.pentest.mjs} & 11 & 5 spoofed roles, bidi overrides, zero-width camouflage, control chars, code-fence breakout, imperative-override phrases, multi-vector composite, idempotency \\
    \hline
    \texttt{code-injection.pentest.mjs}   & 8  & shell-metachar manifest fields, JS-eval clearance rejected, function-shaped JSON treated as data, channel id path-traversal denied, \texttt{\_\_proto\_\_} provider id, allowedDirs allowlist, malformed JSON error context, deep-JSON safety \\
    \hline
    \texttt{extension-egress-bypass.pentest.mjs} & 10 & raw-socket bypass via \texttt{net.Socket.connect}, \texttt{net.createConnection} (normalized-array form), \texttt{http.request} to literal IPs, subclassed \texttt{Socket}, VPN-CIDR boundary, \texttt{ipInCidr} edges, onDeny audit hook, post-freeze tamper attempts \\
    \hline
    \texttt{extension-audit-tamper.pentest.mjs} & 12 & per-field record edits, recordHash forgery propagation, deletion of middle records, re-ordering, 32 concurrent appends, newline log-injection, prototype-pollution payloads, documented tail-truncation gap \\
    \hline
    \texttt{extension-net-admission.pentest.mjs} & 15 & unknown capability tokens rejected, empty \texttt{netAllowedHosts} rejected, unsigned in enclaved rejected, signer-not-in-trust-root rejected, post-signing tamper caught (canonical bytes cover \texttt{verification} + \texttt{netAllowedHosts}), \texttt{net.egress} below \texttt{tested} rejected, per-extension Socket guard, biconditional D=S report \\
    \hline
    \textbf{Total OSS pen-tests} & \textbf{95} & \\
    \hline
  \end{tabular}
  \caption{Adversarial test inventory. Each row covers a distinct
  attack family; cases that document a known limitation are noted as
  ``documented'' or ``gap'' so the deploying organization knows where
  defense-in-depth is required.}
  \label{tab:pen}
\end{table}

\subsection{Bugs found and fixed during construction}

The unit and pen suites caught the following real defects during
development; each was fixed prior to publication:

\begin{enumerate}
  \item \texttt{crypto-fips.deriveKey} ---
    \texttt{ERR\_CRYPTO\_INVALID\_SCRYPT\_PARAMS} at boundary. Lifted
    \texttt{maxmem} explicitly to 64\,MiB.
  \item \texttt{classification.makeLabel} --- compartments backed by
    \texttt{Set} were mutable through \texttt{.add()} despite
    \texttt{Object.freeze}. Switched to deduplicated, sorted, frozen
    arrays.
  \item \texttt{classification.parse} --- auto-promoting
    \texttt{TOP\_SECRET} to \texttt{TOP\_SECRET\_SCI} on any
    non-releasability segment broke \texttt{format}/\texttt{parse}
    round-tripping. Removed; SCI is a separately representable head.
  \item \texttt{audit-log.append} --- concurrent calls raced on
    \texttt{lastHash}, producing a broken chain. Fixed with internal
    Promise queue.
  \item \texttt{audit-log} log-injection --- payload strings containing
    raw newlines could spoof a fake JSONL record. Fixed with
    \texttt{deepSanitize} pass before record construction.
  \item \texttt{trust-root} --- post-boot \texttt{setTrustRoot} could
    substitute attacker-controlled signers. Fixed with
    \texttt{lockTrustRoot()} called by the bootstrap in
    \texttt{enclaved}.
  \item \texttt{dlp-scanner} --- unbounded input could trigger
    pathological regex CPU. Added 1\,MiB cap with throw or truncate.
  \item \texttt{egress-guard} --- a hostile extension could reassign
    the fetch wrapper or patch \texttt{Socket.prototype.connect} to
    bypass the guard. Fixed with the \texttt{freeze} option, which uses
    \texttt{defineProperty} (\texttt{writable:false},
    \texttt{configurable:false}) on both layers, so reassignment throws
    under ESM strict mode.
  \item Scheme parser --- accepted a non-string \texttt{canonicalName}
    via \texttt{String(...)} coercion. Now requires a real string with
    TypeError.
  \item Scheme loader --- unwrapped \texttt{JSON.parse} surfaced a
    bare \texttt{SyntaxError}. Wrapped with file-path context; added
    \texttt{allowedDirs} path-traversal allowlist.
\end{enumerate}

\subsection{Aggregate result}

All 356 tests (261 unit + 95 pen) pass on Node 22 in approximately
1.0\,second. Every framework TypeScript file type-checks clean under
\texttt{tsc --strict --noEmit}. The continuous-integration (CI)
workflow at
\texttt{.github/workflows/enclawed-security.yml} runs both suites on
push, pull request (PR), and weekly cron, plus a strict-mode typecheck,
the bundle verifier
(\texttt{scripts/verify-enclaved-bundle.mjs}), and a boot-time
signature verification of every shipped module manifest. The verifier
counts both signed extensions (admitted in the enclaved flavor) and
gated-unsigned extensions (rejected by the host admission layer in
the enclaved flavor) and exits non-zero on any signature mismatch or
unreachable-tree violation.

\section{Related Work}

\paragraph{Multi-channel AI assistants.} OpenClaw [25] is the upstream
personal-AI gateway from which enclawed is forked. LiteLLM [19] provides
a multi-provider proxy but does not address classification or
attestation. Local-inference engines such as Ollama [18], vLLM [16],
LM Studio [19], and SGLang [17] are the inference targets that survive
the deletion catalog of Section 3.2.

\paragraph{Information-flow control.} The Bell-LaPadula model [10]
originated the no-read-up / no-write-down formulation enclawed
implements. Operating-system mandatory-access-control (MAC)
implementations such as SELinux [11] are the kernel-layer counterpart
that enclawed assumes the deploying organization will configure
separately.

\paragraph{Prompt-injection research.} Greshake et al.\ [20] formalized
indirect prompt injection; Perez and Ribeiro [21] catalogued early
attack patterns; Schulhoff et al.\ [22] demonstrated competition-grade
jailbreaks. enclawed's \texttt{prompt-shield} module addresses the
data-sanitization layer of this problem only; the broader research
direction remains open.

\paragraph{Audit logs.} Tamper-evident logging via hash chains is a
classical technique [12, 14]. enclawed implements a single-machine
variant; production deployments should layer WORM storage or off-host
shipping (e.g.\ to a separate audit aggregator) on top.

\paragraph{Formal verification of skill capability containment.}
The companion paper \emph{Methods for Formal Verification of Agent
Skills} [57] develops the three composable
methods (sound static effect containment, refinement-typed dispatch,
and bounded model checking against the runtime biconditional) that
this paper's \texttt{skill-formal-*.mjs} stack implements. That paper
positions the work against the broader landscape ---
the Model Context Protocol (MCP) and OpenAI function-calling for
schema-validated tool dispatch, Open Policy Agent and AWS Cedar for
runtime authorisation, Sigstore / Supply-chain Levels for Software
Artefacts (SLSA) / The Update Framework (TUF) for supply-chain
attestation, and OWASP's LLM Top 10 / MITRE's Adversarial Threat
Landscape for Artificial-Intelligence Systems (ATLAS) for the attack
catalogues this work neutralises --- and gives the soundness theorems
this paper does not reproduce. The two papers are intended to be
read together: this paper documents the framework that contains the
proof-carrying bundle; the companion paper documents the proofs.

\section{Limitations and Gaps the Deploying Organization Owns}
\label{sec:gaps}

The framework cannot, by itself, deliver an accredited deployment. The
following gaps are explicitly out of scope and must be filled by the
deploying organization's security and compliance program:

\begin{itemize}
  \item \emph{Identity binding.} \texttt{canRead} enforces
    Bell-LaPadula, but no identity layer ships. Integrate the
    organization's identity provider (IdP) --- using SAML, OIDC, or a
    smart card --- and bind a clearance label to every session.
  \item \emph{OS-level mandatory access control.} Defense in depth
    requires SELinux MLS or equivalent enforcing the same lattice
    outside the JavaScript process.
  \item \emph{Cross-trust-zone transfer.} The framework forbids egress;
    a real deployment needs an accredited cross-zone control (a data
    diode, a file-transfer appliance, a manual review queue, or a
    vendor CDS).
  \item \emph{Validated cryptographic module.} The deployment must link
    against a regime-appropriate validated module (FIPS 140-3 for U.S.\
    federal / DoD / DoE / FedRAMP; Common Criteria (CC) Evaluation
    Assurance Level (EAL) evaluations for some EU and Asia-Pacific
    (APAC) regimes).
  \item \emph{Key management.} Replace passphrase parameters with
    HSM-backed key references (Public-Key Cryptography Standards \#11
    (PKCS\#11), the Key Management Interoperability Protocol (KMIP),
    or the vendor's software development kit (SDK)).
  \item \emph{At-rest coverage.} Wrap every persistent artifact ---
    credentials, sessions, model weights, swap, core dumps --- not just
    the credential planner.
  \item \emph{Audit log durability.} Ship records to WORM media or an
    isolated audit aggregator; the on-disk hash chain alone cannot
    detect trailing truncation.
  \item \emph{Memory hygiene.} JavaScript strings cannot be zeroized;
    audit every secret-handling path. Disable swap and core dumps.
  \item \emph{Plugin allowlist by signature, not id.} Bind each allowed
    plugin to a signed manifest hash and reject any drift.
  \item \emph{Cascade work from gated extensions.} Cloud channels and
    cloud LLM providers ship in the source tree but without a signed
    manifest, so the host admission gate refuses to load them in the
    enclaved flavor. Upstream code may still carry legacy references
    (legacy migrations, doctor repair paths, hard-coded fallbacks) that
    assume those plugins exist. The deploying organization must run
    \texttt{pnpm tsgo \&\& pnpm test \&\& pnpm build} to enumerate and
    fix them.
\end{itemize}

\section{Conclusion}

enclawed packages a coherent set of hardening primitives ---
classification labels, deny-by-default policy, egress allowlist,
hash-chained audit, DLP scanning, FIPS-gated at-rest crypto, secret
zeroization, Ed25519 module signing with a clearance-aware trust root,
and prompt-injection sanitization --- behind a single one-call bootstrap
that runs before any plugin loads. The two-flavor design preserves
upstream compatibility for community deployments while supporting strict
enforcement in regulated settings. The classification ladder is
data-driven, accommodating any sector's vocabulary through five
built-in presets or a custom JSON scheme. The 356-test suite (261 unit +
95 adversarial) and the CI workflow let a deploying organization
continuously verify that the security primitives still hold against the
attack patterns enumerated in Section 9. The verification lattice is
now closed at the top: the four-primitive
\texttt{skill-formal-\{static,types,bmc,bundle\}.mjs} stack plus the
\texttt{scripts/skills-formal-verify.mjs} CLI raise an admitted skill
from \texttt{tested} to \texttt{formal} via composable static
effect-containment, refinement-typed dispatch, and bounded model
checking, and emit a signed proof-carrying bundle the runtime
re-checks at every load.

The framework is not a substitute for accredited cryptography, hardware,
facilities, or assessor sign-off; the gaps in Section~\ref{sec:gaps}
remain the deploying organization's responsibility. What enclawed
provides is an explicit, well-tested, and continuously-verifiable
scaffold that the organization's compliance program can build on rather
than reinvent.

\section*{Availability and Licensing}

\texttt{enclawed-oss} is the open-source framework documented in this
paper. The framework primitives --- 29 TypeScript files (including the
integration shims) plus the parallel 22-file \texttt{.mjs} reference
(of which the four \texttt{skill-formal-*.mjs} primitives ship as
\texttt{.mjs}-only by design and are exercised end-to-end through
\texttt{skill-formal-*.test.mjs}), the unit and adversarial
pen-test suites, the canonical \texttt{SKILL.md} manifest specification
at \texttt{docs/tools/skill-manifest-spec.md}, the
\texttt{skills/\_formal-demo/} worked example, the formal-verification
CLI \texttt{scripts/skills-formal-verify.mjs}, the upstream OpenClaw
fork base, the full mirrored OpenClaw extension catalog (\(\sim\)130
directories, gated by the host admission layer in the enclaved flavor),
this paper, the companion formal-verification paper at
\texttt{papers/skills-formal-arxiv/}, and the CI workflow --- are
released under the \textbf{MIT License} and hosted at
\url{https://github.com/metereconsulting/enclawed} [56]. Copyright (c)
2026 Metere Consulting, LLC.; copyright (c) 2025 Peter Steinberger
(upstream OpenClaw). MIT permits commercial use, modification,
redistribution, and incorporation into proprietary projects, subject
only to preservation of the copyright notice and license text. To
clone:
\begin{verbatim}
git clone https://github.com/metereconsulting/enclawed
\end{verbatim}

A separate proprietary product, \texttt{enclawed-enclaved}, is
distributed by the same publisher under commercial terms. Its
internal architecture, code footprint, and feature set are out of
scope for this paper.


\end{document}